\documentclass[]{spie}  
\usepackage[letterpaper, textwidth=6.75in, textheight=8.75in, top=1in]{geometry}
\usepackage{amsmath}
\usepackage{amsfonts}
\usepackage{amssymb}
\usepackage{mathrsfs}
\usepackage{multicol}
\usepackage{multirow} 
\usepackage{subcaption}
\usepackage{tikz}
\usepackage{graphicx}
\usepackage{listings}
\usepackage{graphics}
\usepackage{xcolor}
\usepackage{listings}
\usepackage{algorithm}
\usepackage{algcompatible}
\usepackage{ctable}

\usepackage{makecell}
\usepackage{tabularx}

\usepackage{epstopdf}
\AppendGraphicsExtensions{.tif}

\newcommand{\R}{\mathbb{R}}

\algnewcommand\INPUT{\item[\textbf{Input:}]}%
\algnewcommand\OUTPUT{\item[\textbf{Output:}]}%
\algnewcommand\RETURN{\item[\textbf{Return:}]}
\definecolor{mygray}{gray}{0.6}

\DeclareMathOperator*{\argmin}{argmin}

\newcommand*\circled[2][.5]{\tikz[baseline=(char.base)]{
    \node[shape=circle, draw, inner sep=1pt, 
        minimum height={\f@size*#1},] (char) {\vphantom{WAH1g}#2};}}

\usepackage{amsmath,amsfonts,amssymb}
\usepackage{graphicx}
\usepackage[colorlinks=true, allcolors=blue]{hyperref}

\title{MACE CT Reconstruction for Modular Material Decomposition from Energy Resolving Photon-Counting Data}

\author[a]{Natalie M. Jadue*}
\author[a]{Madhuri Nagare*}
\author[b]{Jonathan S. Maltz}
\author[a]{\\ Gregery T. Buzzard}
\author[a]{Charles A. Bouman}
\affil[a]{Purdue University, West Lafayette, IN, USA 47907}
\affil[b]{GE HealthCare, Waukesha, WI, USA 53188}
\authorinfo{Further author information: * N.M.J. and M.N. contributed equally to this work.  Send correspondence to G.T.B.\\
N.M.J.: nhalavic@purdue.edu, M.N.: mnagare@purdue.edu, \\
G.T.B.: buzzard@purdue.edu, C.A.B.: bouman@purdue.edu, J.S.M.: jonathan.maltz@ge.com}

\begin{document} 
\maketitle

\begin{abstract}
X-ray computed tomography (CT) based on photon counting detectors (PCD) extends standard CT by counting detected photons in multiple energy bins.  
PCD data can be used to increase the contrast-to-noise ratio (CNR), increase spatial resolution, reduce radiation dose, reduce injected contrast dose, and compute a material decomposition using a specified set of basis materials \cite{danielsson}. 
Current commercial and prototype clinical photon counting CT systems utilize PCD-CT reconstruction methods that either reconstruct from each spectral bin separately, or first create an estimate of a material sinogram using a specified set of basis materials and then reconstruct from these material sinograms.  
However, existing methods are not able to utilize simultaneously and in a modular fashion both the measured spectral information and advanced prior models in order to produce a material decomposition.  

We describe an efficient, modular framework for PCD-based CT reconstruction and material decomposition using on Multi-Agent Consensus Equilibrium (MACE).
Our method employs a detector proximal map or agent that uses PCD measurements to update an estimate of the pathlength sinogram. 
We also create a prior agent in the form of a sinogram denoiser that enforces  both physical and empirical knowledge about the material-decomposed sinogram.
The sinogram reconstruction is computed using the MACE algorithm, which finds an equilibrium solution between the two agents, and the final image is reconstructed from the estimated sinogram.
Importantly, the modularity of our method allows the two agents to be designed, implemented, and optimized independently. 
Our results on simulated data show a substantial (450\%) CNR boost vs conventional maximum likelihood reconstruction when applied to a phantom used to evaluate low contrast detectability.
\end{abstract}

\keywords{computed tomography, CT, photon counting detectors, material decomposition}

\section{Introduction}
\label{sec:intro}

Computed tomography (CT) based on photon counting detectors (PCD) \cite{gronberg2022effects, vision-2020, taguchi-energy-sensitive, clinical-prospects, danielsson} extends standard CT by using measurements of photons at multiple energies.  
When coupled with the spatial content inherent in CT, these spectral measurements have the potential to yield clinically relevant improvements to CT images, including increased contrast-to-noise ratio (CNR), increased spatial resolution, reduced radiation and contrast dose, and material decomposition using a specified set of basis materials \cite{danielsson, vision-2020}. 
Some typical use cases for such high-quality reconstructions include identification of small low-contrast lesions in thin slices, temporal bone imaging, and quantitative plaque characterization in cardiac studies. These applications require high spatial resolution and high CNR \cite{clinical-prospects}. Small pixels necessarily detect fewer counts per unit patient radiation dose. To realize the full potential of photon counting CT (PCCT) for increased spatial resolution without increasing radiation dose, algorithmic methods are required to boost CNR. 

Danielsson et al.\cite{danielsson} discuss various reconstruction and material decomposition approaches; we summarize this discussion here and refer the reader to that source for further details. Existing methods for reconstruction from PCD data fall largely into two classes: either (i) creating one reconstruction for each spectral bin \cite{spectral-diffusion, image-recons-dual-energy} or (ii) creating a material decomposition in a specified material basis, either directly in image space \cite{one-step-recon, foygel} or first in sinogram space followed by a reconstruction \cite{alvarez, cramer-rao-roessl}.  
Reconstructing each spectral bin is straightforward and can be achieved using filtered backprojection or with iterative methods, but this bin-wise approach limits the ability to correct for within-bin beam hardening and to exploit the mutual information between bins.

On the other hand, a material decomposition seeks to produce one image for each of a set of basis materials, typically two primary materials (for K-edge imaging with multiple contrast agents, more material basis functions may be required).
A material decomposition in image space takes bin-wise reconstructions and maps these to material images, but this approach has limited ability to correct for beam-hardening \cite{danielsson}.  
For material decomposition in sinogram or projection space, the photon counts across energy bins at a single projection location are mapped to an estimate of pathlength integrals of the material basis components.  
Typically, this uses the maximum likelihood estimator (MLE) or an approximation to it to model the Poisson distribution of photon counts \cite{mle-roessl, mle-dual-energy}. 

Methods to reduce noise in the resulting material pathlength sinograms include the use of a full-spectrum reconstruction as a prior image \cite{denoising-priorimage}, block-matching \cite{denoising-blockmatching}, dictionary methods \cite{denoising-dictionary}, and deep learning methods \cite{denoising-cnn}.  
However, these existing methods are not able to utilize simultaneously and in a modular fashion both the measured spectral information and advanced prior models in order to produce a material decomposition. 

Another method to reduce noise and physical phenomena such as scatter, pile-up, and cross talk \cite{bornefalk-compton, taguchi-crosstalk, vision-2020, cammin-pileup} is to use a statistical iterative reconstruction method \cite{elbarkri-statistical}. However, these methods differ from the proposed method in their optimization techniques \cite{elbarkri-statistical}, their approximations to the forward model \cite{Mechlem}, and their use cases \cite{dual-energy}.

\begin{figure}
    \centering
    \includegraphics[width=5in]{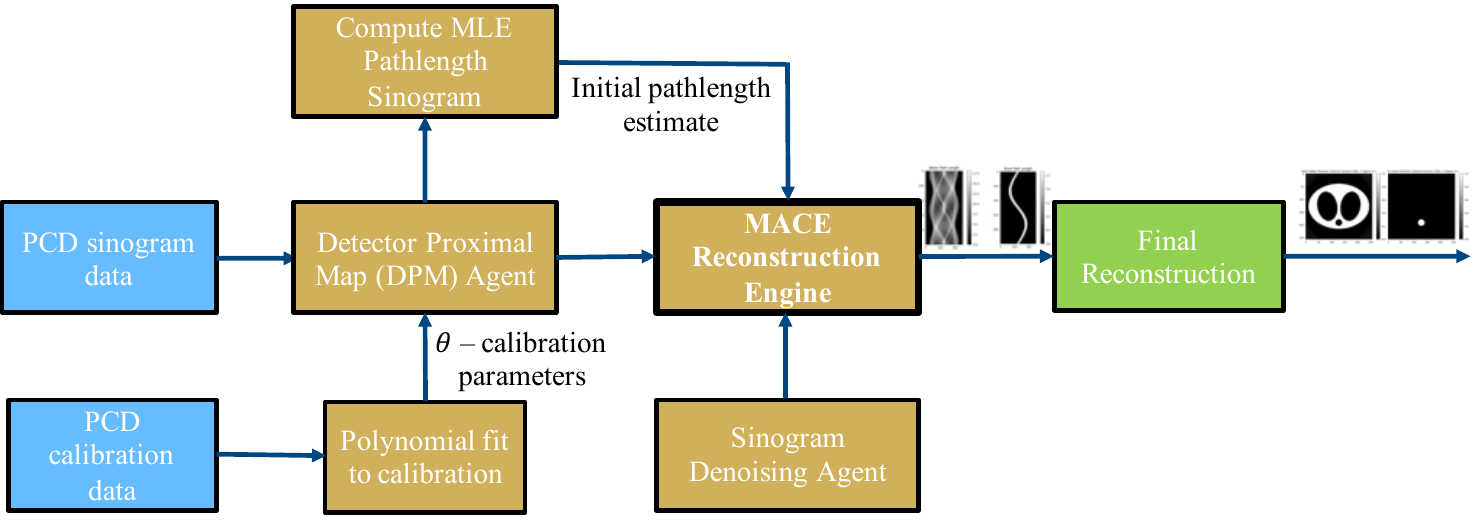}
    \caption{Schematic diagram of algorithm architecture.}
    \label{fig:overview}
\end{figure}

In this paper, we describe an efficient, modular framework for PCD-based CT reconstruction and material decomposition based on the Multi-Agent Consensus Equilibrium (MACE) framework\cite{buzzard2018plug}.
Shown in Figure~\ref{fig:overview}, our method uses a detector proximal map or agent that uses PCD measurements to update an estimate of the pathlength sinogram. 
We also create a prior agent in the form of a sinogram denoiser that enforces  both physical and empirical knowledge about the material-decomposed sinogram.
The sinogram reconstruction uses the MACE algorithm to compute an equilibrium solution between the two agents, and the final image is reconstructed from the estimated sinogram.
Importantly, the modularity of our method allows the two agents to be designed, implemented, and optimized independently. Our experiments with simulated and measured data show that the proposed MACE based reconstruction algorithm outperforms the MLE; on a low-contrast test phantom, the CNR of our method is 4.5 times the CNR of the MLE.

\section{Proposed method}
This paper proposes a modular algorithm for computationally efficient material decomposition using CT measurements obtained with a photon counting detector. 
The algorithm uses the MACE framework\cite{buzzard2018plug} to balance two complementary update operators, or agents:  a data-fitting agent and a prior agent.  The data-fitting agent uses an accurate surrogate function approximation to the detector response to produce an approximate proximal map, which takes a candidate reconstruction as input and produces a reconstruction that better fits the data.  
The prior agent also takes in a candidate reconstruction but then returns a reconstruction with better image properties, typically in the form of reduced noise.  
These two agents are applied sequentially, along with a Lagrangian communication term to enforce a common solution, until convergence.

The MACE algorithm allows for each of these agents to be designed and implemented separately and combined into a computationally efficient and modular reconstruction algorithm. 
With the MACE framework, the proposed algorithm can use material-specific constraints and advanced, pre-trained prior agents, either in the projection domain or in the image domain. 
Here we focus on processing entirely in the projection domain followed by a single reconstruction step to promote computational efficiency, but the modularity of our method means that the use of priors in the image domain is a straightforward variation.

\subsection{Problem Formulation}
\label{sec:problem}

In a Bayesian formulation, our reconstruction problem is to estimate an unknown image $x$ from observed photon counts $y$ using an assumed prior distribution $P(x)$.
Let $y\in \R^{M \times K}$ be the measurements from a photon counting detector with $M$ projections and $K$ energy or spectral bins. 
From $y$, we seek to reconstruct $x \in \R^{N \times L}$ representing the fractional volume of $L$ basis materials in each of the reconstructed $N$ voxels. The maximum a posteriori (MAP) estimate for $x$ is given by
\def\MAP{\mathrm{\footnotesize{MAP}}}
\def\NC{\mathrm{NC}}
\begin{align}
    \hat{x}_{\MAP} 
    & = \argmin_{x\in\Omega}\{-\log P(y|x)-\log P(x)\}, \label{est_map}
\end{align}
where $\Omega$ is the set of possible values for $x$, $P(y|x)$ is the conditional distribution of the observed data $y$ given the unknown image $x$ (or forward model relating $x$ to $y$), and $P(x)$ is the prior model. 

To promote modularity, we incorporate an intermediate pathlength sinogram that links $y$ to $x$.  
More precisely, we let $A \in \R^{M \times N}$ denote the system matrix in distance units (cm), so that the pathlengths of each basis material through the phantom are given by $p=A x$.  
Here, $A_{m,n}$ is the length of intersection of projection ray $m$ with voxel $n$, and $p_{m,l}$ is the distance (in cm) that projection ray $m$ travels through the basis material $l$. 

Using $p=A x$ in \eqref{est_map}, the forward model $P(y|x)$ can be reformulated as a detector forward model, $P_d(y|p),$ which captures details of the detector design by relating measurements $y$ to the pathlength sinogram $p$. Defining $\Omega_p$ as the set of possible values of $p$ and assuming a reconstruction algorithm that reconstructs $x$ from $p$, we obtain the sinogram-based reconstruction approach
\begin{equation} \label{eq:MAP}
    \hat{p}_{\MAP} = \argmin_{p \in \Omega_p}\{f(p) + h(p)\} ,
\end{equation}
where $f(p)= -\log P_d(y|p)$ is the detector forward model and $h(p) = - \log P(p)$ is the prior model in $p$.
The prior model regularizes the estimate, while the detector forward model $P_d(y|p)$ accounts for various physical effects in a photon counting detector, including photon binning, source flux, and detector scatter. 

A shortcoming of the Bayesian approach in  \eqref{eq:MAP} is the need to specify the prior information using a function $h(p)$, which is not available for advanced priors such as neural networks.  To address this shortcoming, we first convert the functions $f$ and $h$ to proximal maps, and then solve \eqref{eq:MAP} using the MACE algorithm\cite{buzzard2018plug}.  The resulting algorithm generalizes immediately to allow the use of advanced priors.  

The first step in applying MACE is to define the detector agent, $F (p;y)$, and the prior agent, $H(p)$, as
\begin{align}
\label{eq:detector_prox}
F (p;y) &= \argmin_q \left\{f(q) + \frac{1}{2 \sigma^2} \|q - p\|^2 \right\} \\
\label{eq:prior_prox}
H(p) &= \argmin_q \left\{h(q) + \frac{1}{2 \sigma^2} \|q - p\|^2 \right\}, 
\end{align}
where $\sigma$ is a parameter that affects the convergence rate.
With these definitions, the MAP estimate of \eqref{eq:MAP} satisfies $\hat{p}_\text{MAP}=p^*$, where 
\begin{equation}  \label{eq:FH}
    F(p^*-u^*;y) = p^* = H(p^*+u^*),
\end{equation}
and $u^*$ is an estimate of the noise removed by the prior agent $H$.  
These equations are called consensus equilibrium equations. 
These equilibrium equations can be solved\cite{mbir_bouman} using Algorithm~\ref{alg:mace}.

Typically, $H(p)$ is a denoising operator that is either image-independent or trained using machine learning methods.
When $H(p)$ is defined directly rather than as a proximal map for $h(p)$, then the MACE reconstruction may not be the solution to any optimization problem.
In our experiments, we choose $H(p)$ to be a linear, space invariant filter with a Gaussian point spread function (psf).

\begin{algorithm}[t]
    \caption{Basic MACE Algorithm}
    \label{alg:mace}
  \begin{algorithmic}[1]
    \INPUT $y$ (data), $p_{\mathrm{init}}$ (initial reconstruction), $\rho$ (step size), $H$ (denoiser), $F$ (data-fitting update)
    \OUTPUT $p$ (final reconstruction in material path lengths) 
    \STATE $p \leftarrow p_{\mathrm{init}}$
    \FOR{$i=0: N_{\mathrm{MACE}}-1$}
        \STATE $p_1 \leftarrow 2H(p)-p$
        \STATE $p' \leftarrow F (p_1; y )$
        \STATE $p_1 \leftarrow 2p'-p_1$
        \STATE $p \leftarrow (1-\rho)p+\rho p_1$
    \ENDFOR
\RETURN $p'$ 
  \end{algorithmic}
\end{algorithm}


\subsection{Photon Counting Detector Model}
\label{sec:forward}

In this section, we derive a simplified model for the detector log likelihood, $f(p) = -\log P_d (y|p)$, based on the properties of a photon counting detector.
In Sections~\ref{sec:calibration} and~\ref{sec:surrogate}, we explain how the model can be calibrated and efficiently computed using a quadratic surrogate function for $f$ that allows for fast approximation of $F(p;y)$.

To define $f$, let $y_j\in \R^K$ be the vector of photon counts for each of $K$ energy bins for the $j^{th}$ projection.
We assume that $y_j$ is composed of independent Poisson distributed random variables with mean $\lambda_j (p) \in \R^K$, where $p\in \R^L$ is a vector of path lengths for each of $L$ basis materials.
So, for each detector we assume that the expected number of photon counts, $\lambda_j (p)$, is completely determined by the pathlength through the object being imaged, and that the distribution of those photon counts are i.i.d.\ and Poisson.
Note that this is an approximation since the expected photon count can vary due to effects such as scatter, and that the distribution of the counts can vary from i.i.d.\ Poisson due to effects such as Compton scatter in the detector\cite{gronberg2022effects}, pile-up, and other nonlinear interactions.

In practice, we are not usually given the raw photon counts $y_{j,k}$, since these photon counts are typically first normalized to the total photon count in air.
Consequently, our measurement for the $j^{th}$ projection is given by the vector $T_j = [ T_{j,0} , \cdots , T_{j,K-1} ]\in \R^K$ where
\begin{align}
\label{eq:TransmissionMeasurements}
T_j = \frac{y_j }{\lambda_{j, \Sigma}}
\end{align}
and $\lambda_{j, \Sigma} = \sum_{k=1}^{K} \lambda_{j,k} (0)$ is a scalar representing the total photon count for the $j^{th}$ projection for air (or when there is no phantom present).
If we assume the same photon count for air for each view angle, then the quantity $\lambda_{j, \Sigma}$ can be obtained by performing a calibration ``blank scan'' with all objects removed from the scanner and summing over all energy bins for each detector.

We also define a corresponding detector response function (DRF) defined by $\phi_j = [ \phi_{j,0} , \cdots , \phi_{j,K-1} ]\in \R^K$ where 
\begin{align}
\label{eq:detector_response_function}
\phi_j (p ) 
    = -\log \frac{ \lambda_j (p ) }{ \lambda_{j, \Sigma} }
    = -\log \frac{ E [ y_j | p_j = p ] }{ \lambda_{j, \Sigma} } \ ,
\end{align}
where we will assume that the function $\phi_j (p)$ has been precisely measured in a calibration process for each of the scanner's detectors.

The empirically measured DRF function $\phi$ accounts for a wide range of systematic effects that can change the observed photon count as a function of material pathlength.  This includes primary effects such as attenuation through the material, as well as secondary effects such as beam-hardening, photon pile-up in the detector, and Compton scatter in the detector that causes a single photon to be detected as two or more lower energy events.  However, this does not capture effects that result from scatter in the tissue sample or other nonlocal effects.  

Using these definitions along with the form of the Poisson distribution, we can derive an expression for the negative log likelihood as:
\begin{align*}
f_j (p) &= \sum_{k=0}^{K-1} - \log P_d ( y_{j,k} | p ) \\
    &= \sum_{k=0}^{K-1} \lambda_{j,k}(p) - y_{j,k} \log \lambda_{j,k}(p) + C_j (y) \\
    &= \lambda_{j, \Sigma} \sum_{k=0}^{K-1} \bigg[\frac{\lambda_{j,k}(p)}{\lambda_{j, \Sigma}}- \frac{y_{j,k}}{\lambda_{j, \Sigma}} \log(\lambda_{j,k}(p)) \bigg]+C_j (y) \\
    &= \lambda_{j, \Sigma} \sum_{k=0}^{K-1} \big[e^{-\phi_{j,k}(p)}+ T_{j,k} \, \phi_{j,k}(p)\big] + C_j^\prime (y) \ .
\end{align*}
Dropping the constant, we can write the loss function for the $j^{th}$ detector as 
\begin{align}
\label{eq:SingleChannelDetectorResponse}
f_j (p) = \lambda_{j, \Sigma} \big[e^{-\phi_j (p)}+ T_j \circ \phi_j (p)\big] \mathbf{1} \ ,
\end{align}
where $\mathbf{1}$ represents a column vector of $1's$, $\circ$ represents the Hadamard product of the vectors, and the exponential function is assumed to be applied point-wise.

\subsection{Detector Calibration}
\label{sec:calibration}

To model non-linear effects such as beam hardening and scatter, we represent the DRF for the $j^{th}$ detector and $k^{th}$ energy bin using a low-degree polynomial (order $P=4$ in our experiments) denoted by
\begin{align*}
\phi_j (p) = \phi (p; \theta_j ) \ ,
\end{align*}
where $\phi (p,\theta_j)$ is a set of $K$ polynomial functions of $p \in \R^L$ with coefficients $\theta_j$.
More specifically, for $L=2$ materials, we have that
$$
\phi (p; \theta_j ) = \sum_{a=0}^{P} \sum_{b=0}^{P} \, \theta_{j,a,b} \, p_0^a \, p_1^b \ .
$$
where $\theta_j$ is a parameter vector with $K\times (P+1)^2$ components. 
Note that coefficients $\theta_j$ are determined for each detector separately,
and that $\theta_{j,a,b}\in \R^K$ is a vector of coefficients for each energy bin.

We note that for photon counting detectors with relatively narrow energy bins, $\phi_{j,k} (p)$ is approximately a linear function of $p$ plus a constant offset, so the low order polynomial is appropriate,
and we include a constant term since the value of $\phi_{j,k} (0)$ from~\eqref{eq:detector_response_function} is not necessarily 0.

In order to estimate the coefficients of the polynomial $\theta_{j,k}$, we first select a set of path lengths $\{ p_s \}_{s=1}^{S}$.
Then for path length, $p_s$, and for each detector and bin $(j,k)$, we make repeated measurements of the photon counts.
These repeated measurements are then averaged to form an estimate of the expected photon count, $\hat{\lambda}_j (p_s )$.
From this average, we can then compute an estimate of the DRF given by
$$
\hat{\phi}_j (p_s) = - \log \frac{ \hat{\lambda}_j (p_s ) }{ \hat{\lambda}_{j,\Sigma } } \ .
$$
Then from these estimates of the DRF, we select the polynomial coefficients for each detector, $\theta_{j}$, based on a least squares fit of the polynomial to the observed points as follows.
\begin{align}\label{eq:thetamin}
    \hat{\theta}_{j} = \argmin_{\theta_j}\sum_{p\in{p_s}} \|\phi(p;\theta_j) - \hat{\phi}_j (p)\|^2,
\end{align}
More generally, we could use another approach to fitting such as a weighted least squares or minimum cross entropy.

A typical set of points used for calibration spans material densities seen in clinical applications for the materials used in calibration.
For example, if the calibration is done using the material bases of polyethylene (PE) and polyvinyl chloride (PVC), we use various points spanning 0 to 40 cm of PE and 0 to 5 cm of PVC.
These points are dispersed through out the 2D space so that an accurate polynomial fit can be computed for each detector and energy bin.

\subsection{Fast Computation of Detector Proximal Map}
\label{sec:surrogate}

In this section, we describe how the proximal map of~\eqref{eq:detector_prox} is efficiently computed.
First, we note that since the terms are decoupled for each detector, we can write the proximal map solution in the separable form of
\begin{align*}
F (p) = \left[ F_0 (p_0 ) , \cdots , F_{M-1} (p_{M-1}) \right] \ ,
\end{align*}
where
\begin{align}
\label{eq:SingleChannelDetectorProximalMap}
F_j (p_j ) &= \argmin_{q\in \R^K} \left\{ f_j (q ) + \frac{1}{2 \sigma^2} \|q - p_j \|^2 \right\} \ ,
\end{align}
and $f_j (q)$ is defined in~\eqref{eq:SingleChannelDetectorResponse}.
Importantly, the detector proximal map can be computed at each detector in parallel.

In order to make the computation of the single channel detector proximal map of \eqref{eq:SingleChannelDetectorProximalMap} both fast and accurate, we use a majorization approach based on a quadratic surrogate function\cite{mbir_bouman}.
The exact proximal map can then be computed by iterating the surrogate update.

First, we notice that the we can express the DRF in the form
\begin{align}
\label{eq:f_j(phi)}
f_j (q ) = \lambda_{j,\Sigma} \; g( \phi_j (q) ; T_j ) \, \mathbf{1} \ ,
\end{align}
where the function $g$ is assumed to use row vectors for input and output, and $g$ is given by
\begin{align}
\label{eq:g(z)}
g(z ; T ) = e^{-z} + T \circ z \ .
\end{align}

\begin{algorithm}[t]
    \caption{ Partial Update Detector Proximal Map - Single Detector} 
    \label{alg:Detector_Prox_Map}
  \begin{algorithmic}[1]
    \INPUT $p$, $p'$,  $T$, $\lambda_\Sigma$, $\theta$, $\sigma$, $N$
    \OUTPUT $F(p; p')$ updated vector of path lengths 
    \FOR{$i=0: N-1$}
      \STATE $z_{\min} \leftarrow \phi(p^\prime ; \theta )-\varepsilon$
      \STATE $A \leftarrow \nabla \phi(p^\prime ; \theta )$
      \STATE $b\leftarrow - e^{-\phi(p')}+ T  $
      \STATE $c \leftarrow 
      2 \frac{e^{-z_{\min}}-e^{-\phi(p'; \theta )} \circ (1+\phi(p'; \theta )-z_{\min})}{(\phi(p'; \theta)-z_{\min})^2}$
      \STATE $C \leftarrow \text{diag}\{c\}$
      \STATE $\alpha \leftarrow \sigma \sqrt{ \lambda_\Sigma }  $      
      \STATE $H \leftarrow A^t CA+\frac{1}{\alpha^2}I$
      \STATE $q \leftarrow A^t (CAp'-b)+\frac{1}{\alpha^2}p$
      \STATE $p' \leftarrow H^{-1}q$ 
    \ENDFOR
    \RETURN $p'$
  \end{algorithmic}
\end{algorithm}

Using Theorem 17.1\cite{mbir_bouman}, the optimal quadratic surrogate for the function $g( z ; T) $ over the interval $[z_{min}, \infty)$ with base point $z = z'$ is given by
\begin{align}\label{surrr_g}
    g(z; z' , T ) = b^t \, (z-z') + (z-z')^t \, \frac{ C }{2} \, (z-z')
\end{align} 
where 
\begin{align*}
b &= -e^{-z'} + T  \\
C &= \mbox{diag} \left\{ 2 \frac{g(z_{\min}; T)-g(z'; T) + g'(z'; T) \circ (z'-z_{\min}) }{ (z'-z_{\min})^2} \right\}  \ ,
\end{align*}
where the division is point-wise, and we assume that $z> z_{\min}$ in each component.
With the given assumptions, $\phi(p)>0$, so we could take $z_{\min}$ to be 0 for purposes of approximating \eqref{eq:f_j(phi)}.
However, empirically we have found that this estimate is too conservative and results in slow convergence. 
Since $z'$ is already an estimate of the minimum, we use $z_{\min} =z'-\varepsilon$ for small $\varepsilon$ (in our experiments, we use $\varepsilon= 10^{-3}$).  Minimization with this surrogate is essentially an approximate Newton step and so gives faster convergence. 

Using this surrogate function, we can compute a surrogate function for $f_j (q)$ using a Taylor series approximation of the DRF.  This gives
\begin{align*}
f_j ( q ; p^\prime ) 
    &= \lambda_{j,\Sigma} \; g(\phi_j (q); \phi_j (p^\prime ) , T_j ) \, \mathbf{1} \\
    &= \lambda_{j,\Sigma} \left[ b_j^t (\phi_j (q)-\phi_j (p'))  + \frac{1}{2}(\phi_j (q)-\phi_j (p'))^t C_j (\phi_j (q)-\phi_j (p')) \right] \\
    &\approx \lambda_{j,\Sigma} \left[ b_j^t A_j (q - p') + \frac{1}{2} (q - p')^t A_j^t C_j A_j (q - p') \right] \ ,
\end{align*}
where 
\begin{align*}
A_j &= \nabla \phi_j (p^\prime ) \\
b_j &=  - e^{-\phi_j (p^\prime )}+ T_j  \\
C_j &= \mbox{diag} \left\{ 2 \frac{e^{-z_{\min}}-e^{-\phi_j (p')} \circ (1+\phi_j (p')-z_{min})}{(\phi_j (p') - z_{\min})^2} \right\}  \ ,
\end{align*}
and $z_{\min} = \phi(p^\prime ; \theta )-\varepsilon$.

From this, we can compute one partial update of the proximal map starting at $p^\prime$ by introducing  $\alpha = \sigma \sqrt{\lambda_{j,\Sigma}} $ and solving the optimization 
\begin{align*}
F_j (p; p^\prime ) &= \arg \min_{q} \left\{ \lambda_{j,\Sigma} \left[ b_j^t A_j (q - p') + \frac{1}{2} (q - p')^t A_j^t C_j A_j (q - p')  \right] + \frac{1}{2\sigma^2} \| q - p \|^2 \right\} \\
&= \arg \min_{q} \left\{ b_j^t A_j (q - p') + \frac{1}{2} (q - p')^t A_j^t C_j A_j (q - p') + \frac{1}{2\alpha^2} \| q - p \|^2 \right\} \\
&=  \left( A_j^t C_j A_j + \frac{1}{\alpha^2 } I \right)^{-1} \left[ A_j^t ( C_j A_j p^\prime - b_j ) + \frac{1}{\alpha^2 } p \right]  \ .
\end{align*}

Algorithm~\ref{alg:Detector_Prox_Map} gives pseudo-code for computing $F_j ( p ; p^\prime )$ using $N$ partial updates.  We suppress the projection index $j$ for clarity.  
Notice that for a given projection, the function $F ( p ; p^\prime )$ also depends on 
the parameters that are specific to the detector corresponding to the projection (the transmission values, $T$, 
the total photon count, $\lambda_\Sigma$, and
the calibration parameters of the detector, $\theta$), 
along with parameters that are common to all the detectors (the proximal map parameter, $\sigma$, and the number of iterations, $N$).

We will refer to $F_j ( p ; p^\prime )$ as a partial update proximal map because the algorithm approximates the true proximal map as the number of iterations increases.
However, in practice, we would like to use $N=1$ iterations in order to reduce the computation.
We can do this by slightly modifying the MACE algorithm of~\ref{alg:mace} in order to provide a better initial value, $p_{\mathrm{init}}$, for each iteration of the MACE loop.
We note that for this MACE algorithm the prior agent, $H(p)$ may be encoded implicitly as a denoiser or other image enhancement algorithm, including a neural network denoiser.
In this case, we use MACE\cite{buzzard2018plug}, Algorithm~\ref{alg:mace}, which reconciles multiple agents to achieve a consensus solution, $\hat{p}_{\MAP}$. 
The Algorithm~\ref{alg:mace} uses the Mann iteration to solve the MACE equations with a parameter $\rho \in (0, 1)$ that we typically set to $0.8$.

In Algorithm~\ref{alg:mace}, we initialize $p$ with the MLE estimate obtained with Algorithm~\ref{alg:PracticalMLEAlgorithm}. 
In the initial step of Algorithm~\ref{alg:PracticalMLEAlgorithm}, a grid search is performed for each projection.
Algorithm~\ref{alg:PracticalMLEAlgorithm} uses the partial update proximal map to compute the maximum likelihood estimate of the path length for each detector.
In this case, we typically use large values for $\sigma$ and $N_{MLE}$ to ensure convergence to an accurate estimate of the MLE.

Note that due inaccuracies in our approximation of $F$, Algorithm~\ref{alg:PracticalMLEAlgorithm} might diverge for some points. In such cases, we employ Algorithm~\ref{alg:mace}, with operator $H$ that is set to clip $p$ so that it lies in the limits of the material space used in calibration.

\begin{algorithm}[t]
    \caption{MLE Algorithm}
    \label{alg:PracticalMLEAlgorithm}
  \begin{algorithmic}[1]
    \INPUT $T$, $\lambda_\Sigma$, $\theta$, $N_{\mathrm{MLE}}$, $\sigma$, $N_{sub}=1$
    \OUTPUT $p$ the material path lengths 
    \FOR{$j=0: M-1$}
        \STATE $p_j \leftarrow \textrm{GridSearch}(\argmin_p \{ \sum_k e^{-\phi_{j,k}(p; \theta_j )}+\phi_{j,k}(p; \theta_j ) T_{j,k}\}$)
    \ENDFOR
    \FOR{$i=0: N_{\mathrm{MLE}}-1$}
        \STATE $p \leftarrow F(p; p, T, \lambda_\Sigma, \theta, \sigma, N_{sub})$
    \ENDFOR
\RETURN $p$
  \end{algorithmic}
\end{algorithm}

Once the MACE or MLE estimates are computed, we then reconstruct the final image using 3D FDK\cite{fdk} reconstruction on the components of $p$.

\section{Methods}
\label{meth}

We evaluated the proposed algorithm with simulated and measured photon counting data. The simulated data is collected using the GE HealthCare's CatSim software\cite{CatSim}. 
The measured data have been collected and processed by GE HealthCare. 
We utilize 8-energy bin data from two detector rows spanning a 50~cm-diameter field of view. The effects of object scatter, charge sharing, and crosstalk among detectors are not considered in the simulation.

In our experiments, we consider two basis materials owing to the two significant interactions of x-rays used in clinical CT with matter (photoelectric and Compton effects)\cite{textbookJiang, dual-energy}. Polyethylene (PE) and polyvinyl chloride (PVC) are selected as basis materials for the ease of data collection for calibration. The proposed algorithm produces fractional volume maps for the two basis materials. Using the estimated maps, we compute a virtual mono-energy reconstruction at 70~keV.
 Using the volume fractions estimated for PVC and PE, a virtual mono-energy reconstruction at energy $E$ keV is computed  using
\begin{align*}
    \hat x ^ {\textrm{(joint)}}_E = \mu_1(E)\,\hat{x}_{*,1}+ \mu_2(E)\,\hat{x}_{*,2}, 
\end{align*}
where $\mu_l(E)$ is the linear attenuation coefficient for the $l^{th}$ basis material at energy $E$ keV.
If needed, the estimated maps are converted to report fraction volume maps for water and bone using a linear transformation. 

\subsection{Calibration}\label{caldata}
To calibrate the DRF $\phi (p; \theta )$, various slabs of PE and PVC of thicknesses are scanned sequentially for a single view (x-ray source at $0^{\circ}$) multiple times. 
These slabs are kept perpendicular to the iso-ray (line connecting the source and isocenter) and cover the entire detector.
Note that pathlengths at off-center detector channels are adjusted to account for the scanner geometry. 

\subsection{Datasets}
For experiments with simulated data, we use a phantom with inserts of densities close to water to evaluate performance on low-contrast targets. 
We scanned the phantom $12$ times at 120~kVp, 400~mA at 1~s per rotation for 1000 views and averaged the reconstruction results. The scan is reconstructed for the image size of $512 \times 512$ and display-field-of-view (DFOV) of 25.6~cm.

For experiments with measured data, we used data obtained from scans of the Gammex Multienergy phantom, without the body ring. This phantom  contains iodine inserts with various densities. The DFOV is 40~cm which is then cropped to a size of $512 \times 512$ to show just the relevant region.

\section{Results and Discussion}
\begin{figure}[t]
    \centering
    \includegraphics[width=0.46\linewidth]{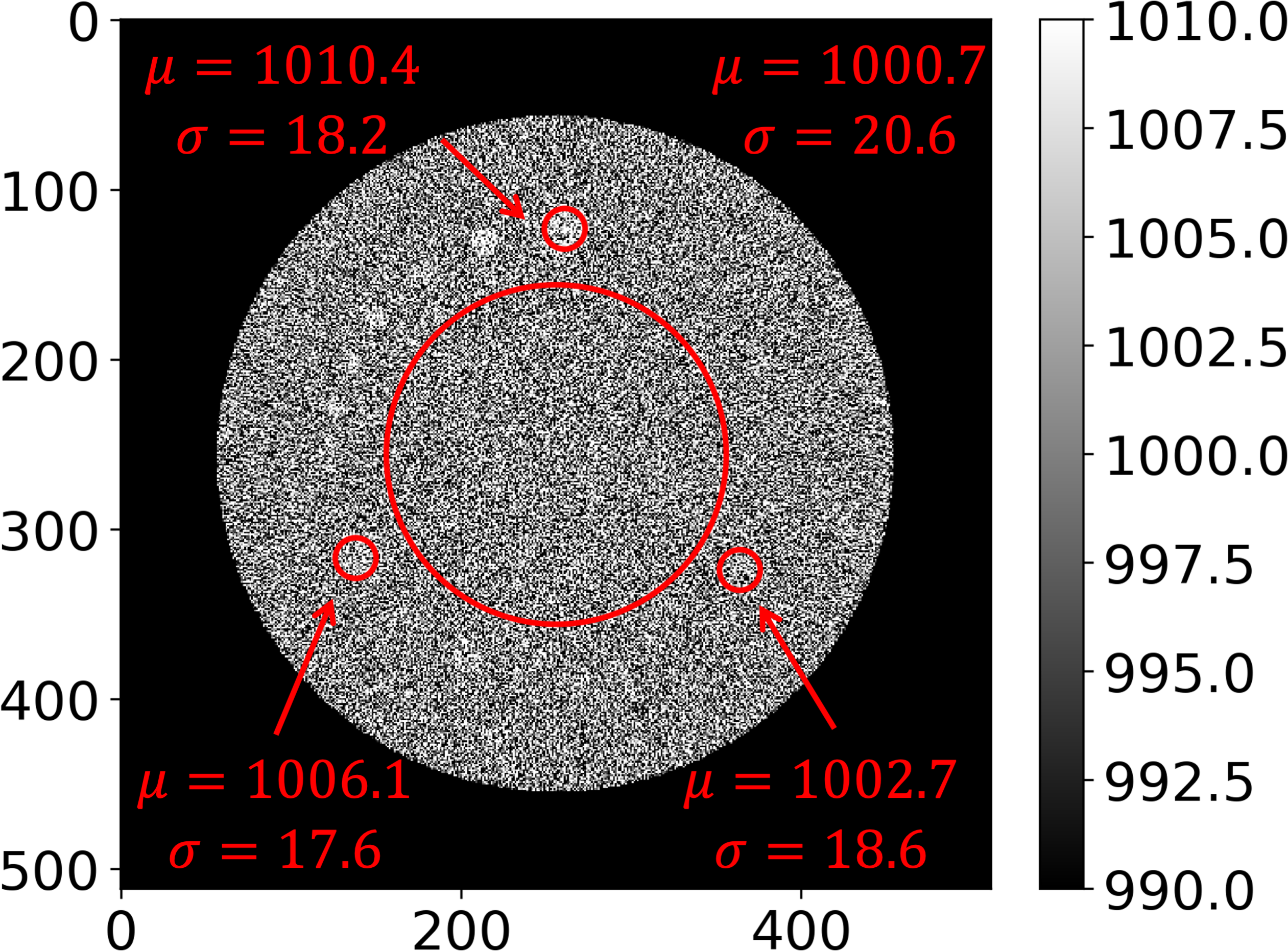}\hfill
    \includegraphics[width=0.46\linewidth]{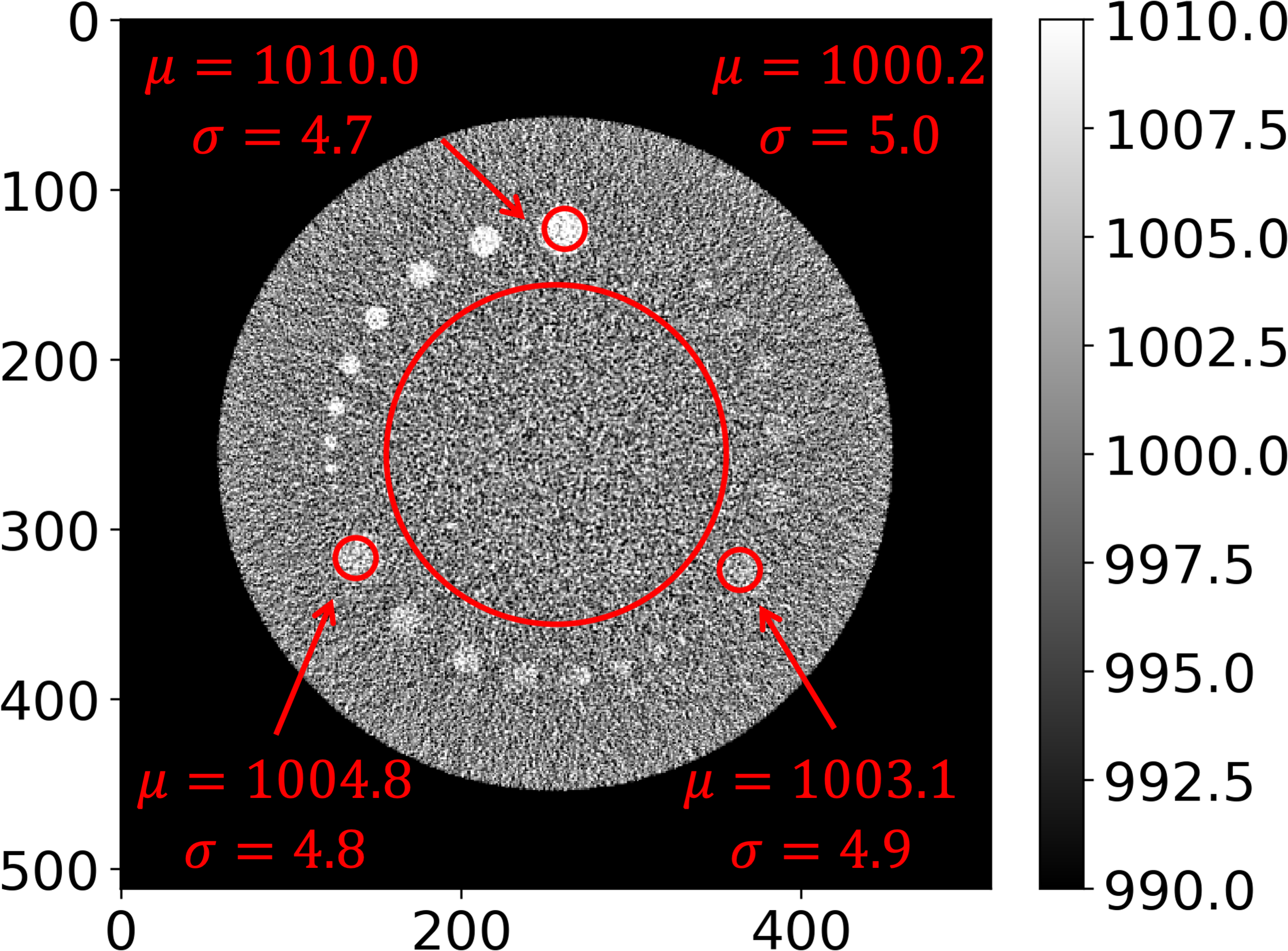}
    \caption{Reconstruction of simulated water background (density 1.0) with inserts of densities 1.01, 1.005, and 1.003.  The results are displayed in Hounsfield units plus 1000, so that water is 1000, air is 0.  Left: Reconstruction from MLE pathlength sinogram.  Right: Reconstruction from MACE-denoised sinogram showing increased CNR (0.5 vs 2.0) and reduced noise. These reconstructions are formed as a linear combination of material decomposed reconstructions, weighted to simulate a 70~keV monoenergetic scan.}
    \label{fig:water_recon}
\end{figure}

The left and right panels of Figure~\ref{fig:water_recon} show the baseline MLE and proposed algorithm reconstructions of the simulated data, respectively. The reconstruction is shown in Hounsfield units (HU), modified so that air is 0 and water is 1000, with a center of 1000~HU and a window width of 20~HU. The result using our proposed method uses a simple linear filter as a prior but is still able to distinguish inserts with less than $0.3\%$ deviation from water density, even for very small feature lengths.  

\begin{figure*}[t!]
\centering
\begin{minipage}{\textwidth} 
\centering
\begin{tabular}{c c c c}
& Water & Iodine & 70 keV\\
\rotatebox[origin=c]{90}{MLE} & 
    \begin{subfigure}[c]{4.5cm}
         \centerline{\includegraphics[width=4.5cm]{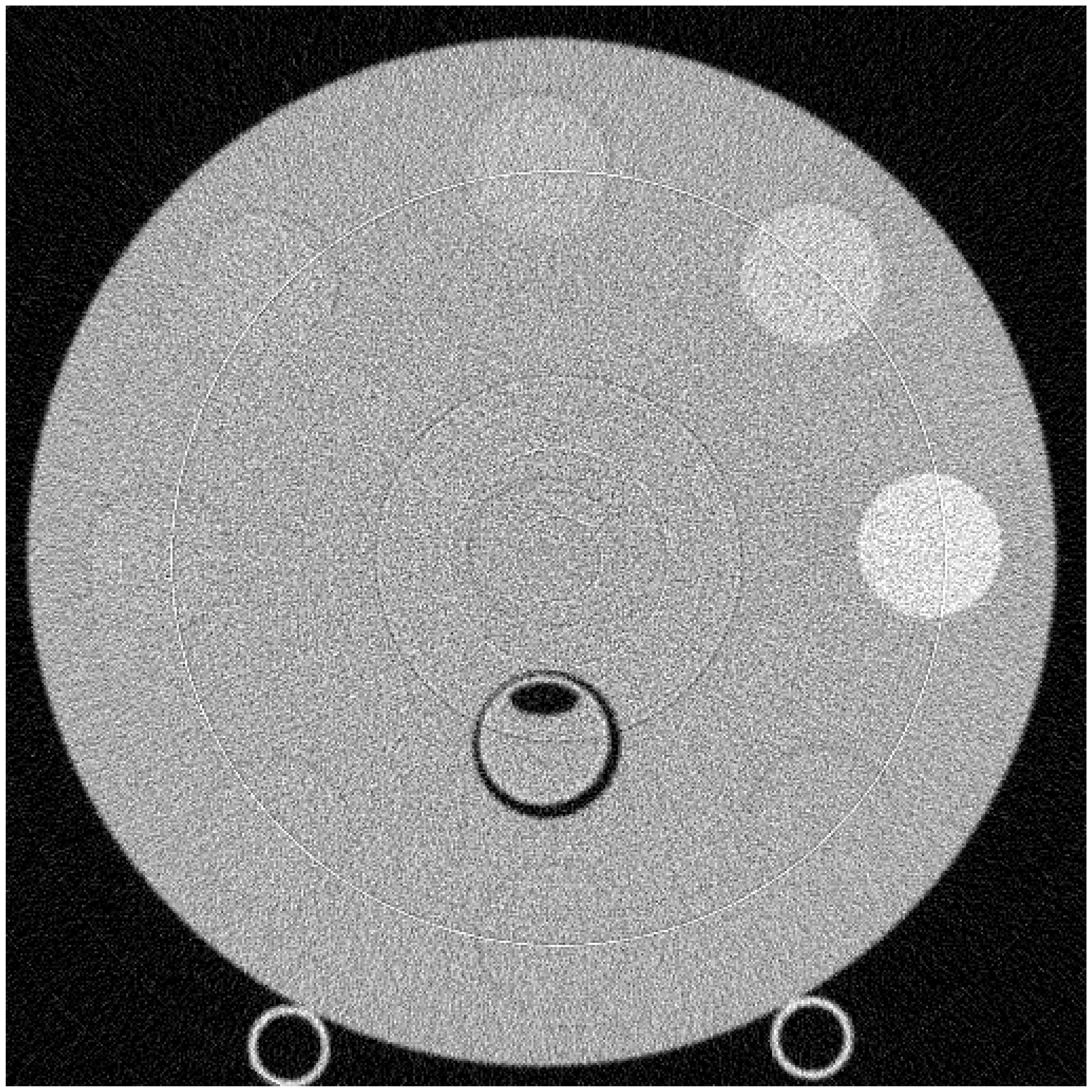}}
        \subcaption{}
        \label{fig:small_gamm_water_mle}
    \end{subfigure}
     &
    \begin{subfigure}[c]{4.5cm}
        \centerline{\includegraphics[width=4.5cm]{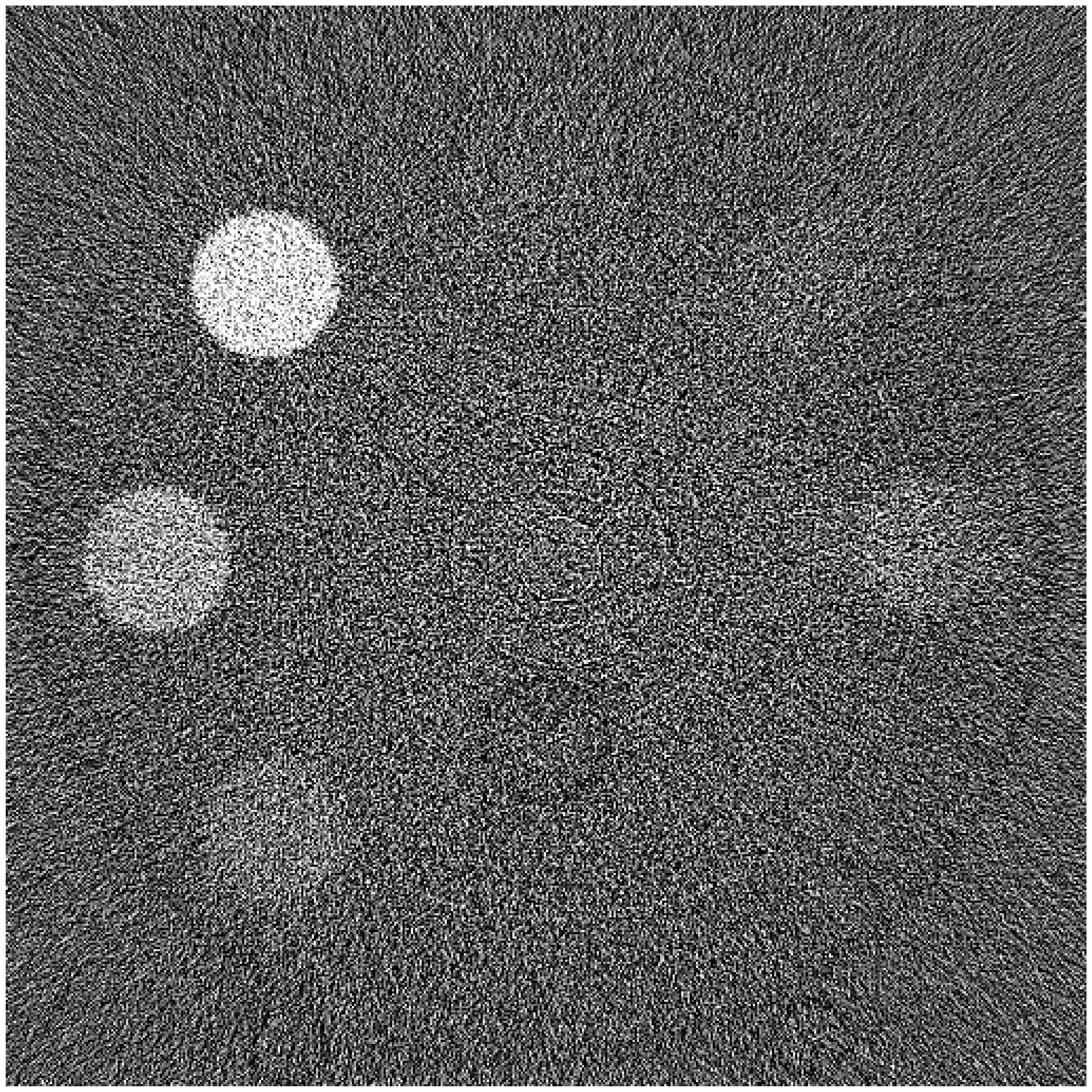}}
        \subcaption{}
        \label{fig:small_gamm_iodine_mle}
    \end{subfigure}
    &
     \begin{subfigure}[c]{4.5cm}
         \centerline{\includegraphics[width=4.5cm]{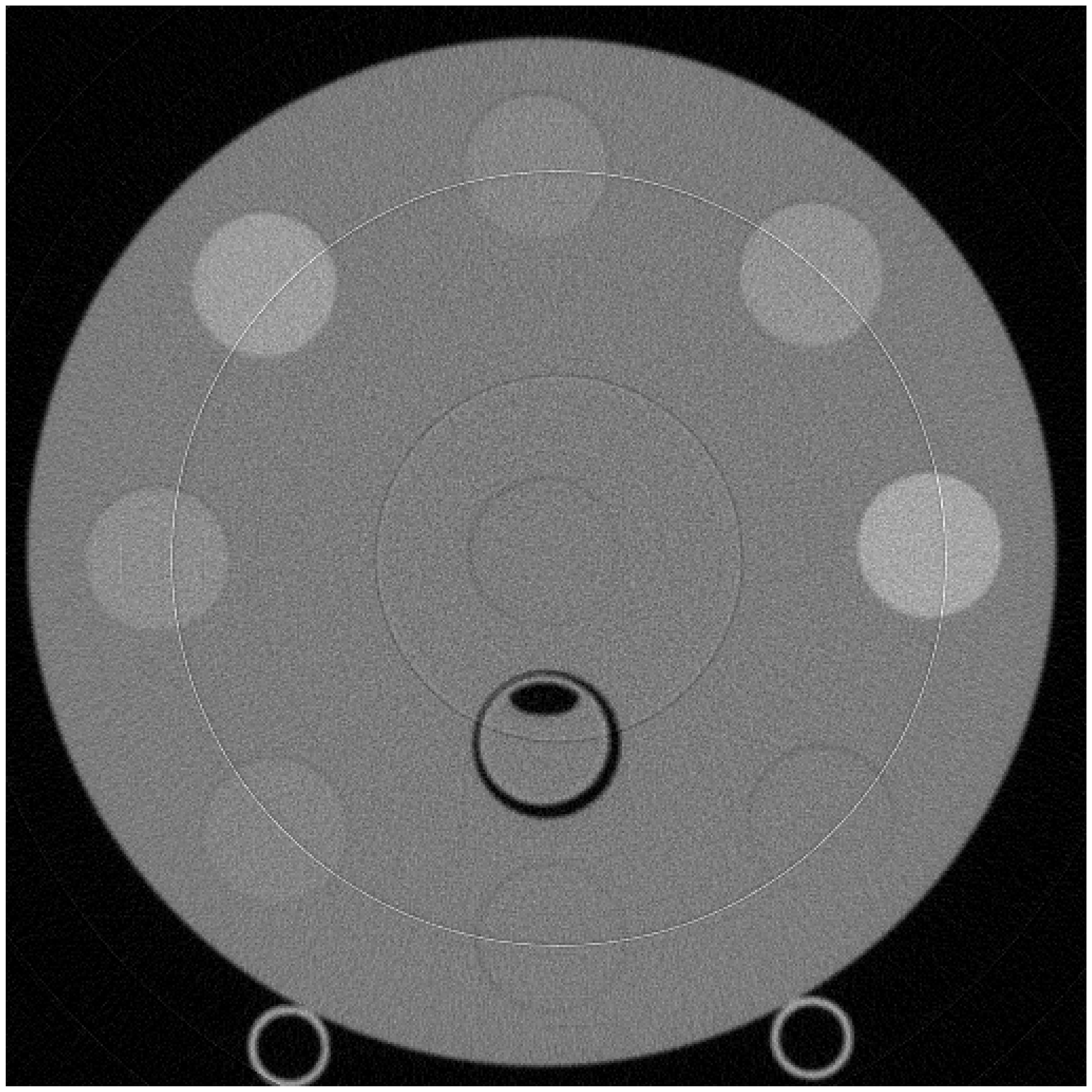}}
         \subcaption{}
         \label{fig:small_gamm_kev_mle}
     \end{subfigure}
\\
\rotatebox[origin=c]{90}{MACE} &
    \begin{subfigure}[c]{4.5cm}
        \centerline{\includegraphics[width=4.5cm]{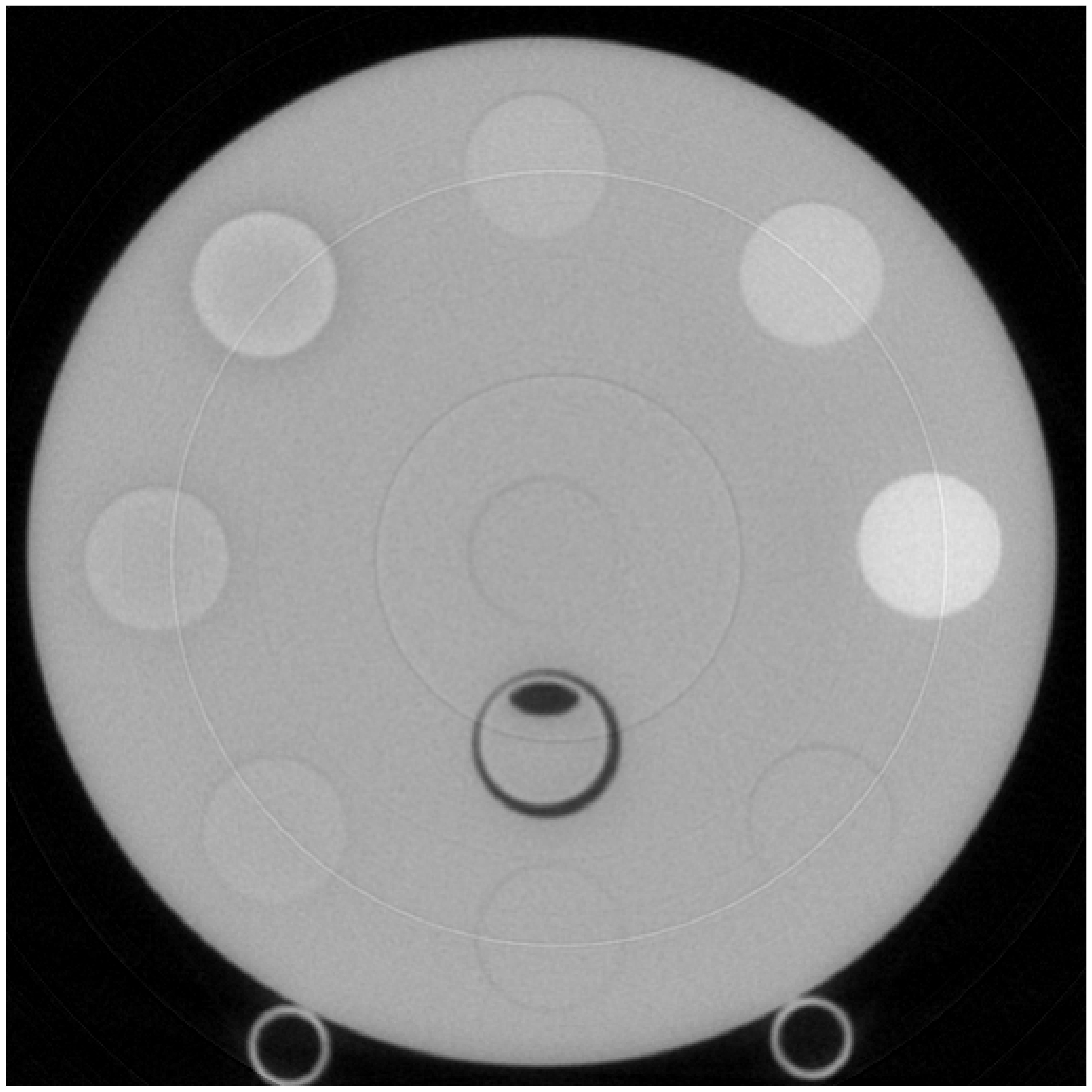}}
        \subcaption{}
        \label{fig:small_gamm_water_new}
    \end{subfigure}
     &
    \begin{subfigure}[c]{4.5cm}
        \centerline{\includegraphics[width=4.5cm]{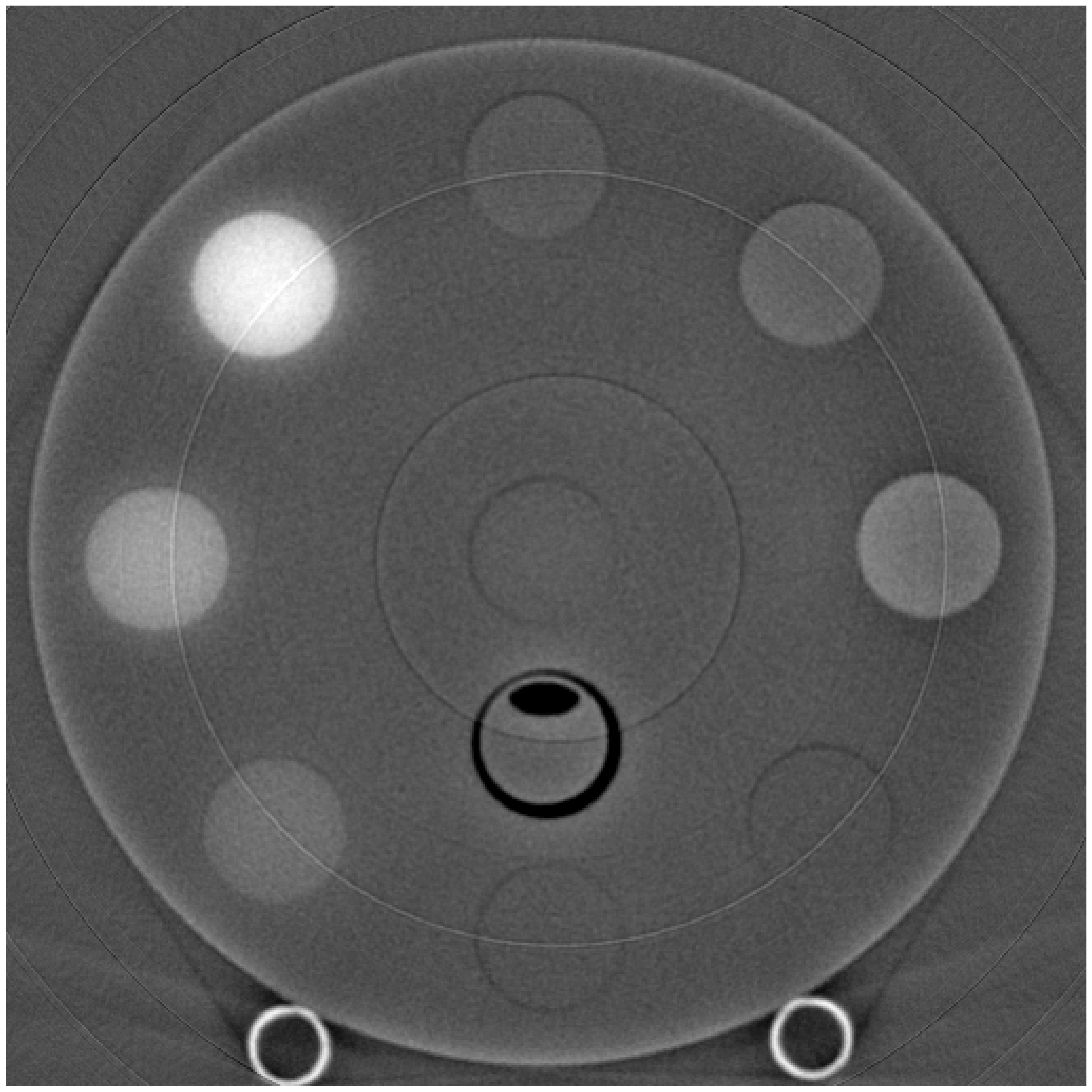}}
        \subcaption{}
        \label{fig:small_gamm_iodine_new}
    \end{subfigure}
    &
     \begin{subfigure}[c]{4.5cm}
         \centerline{\includegraphics[width=4.5cm]{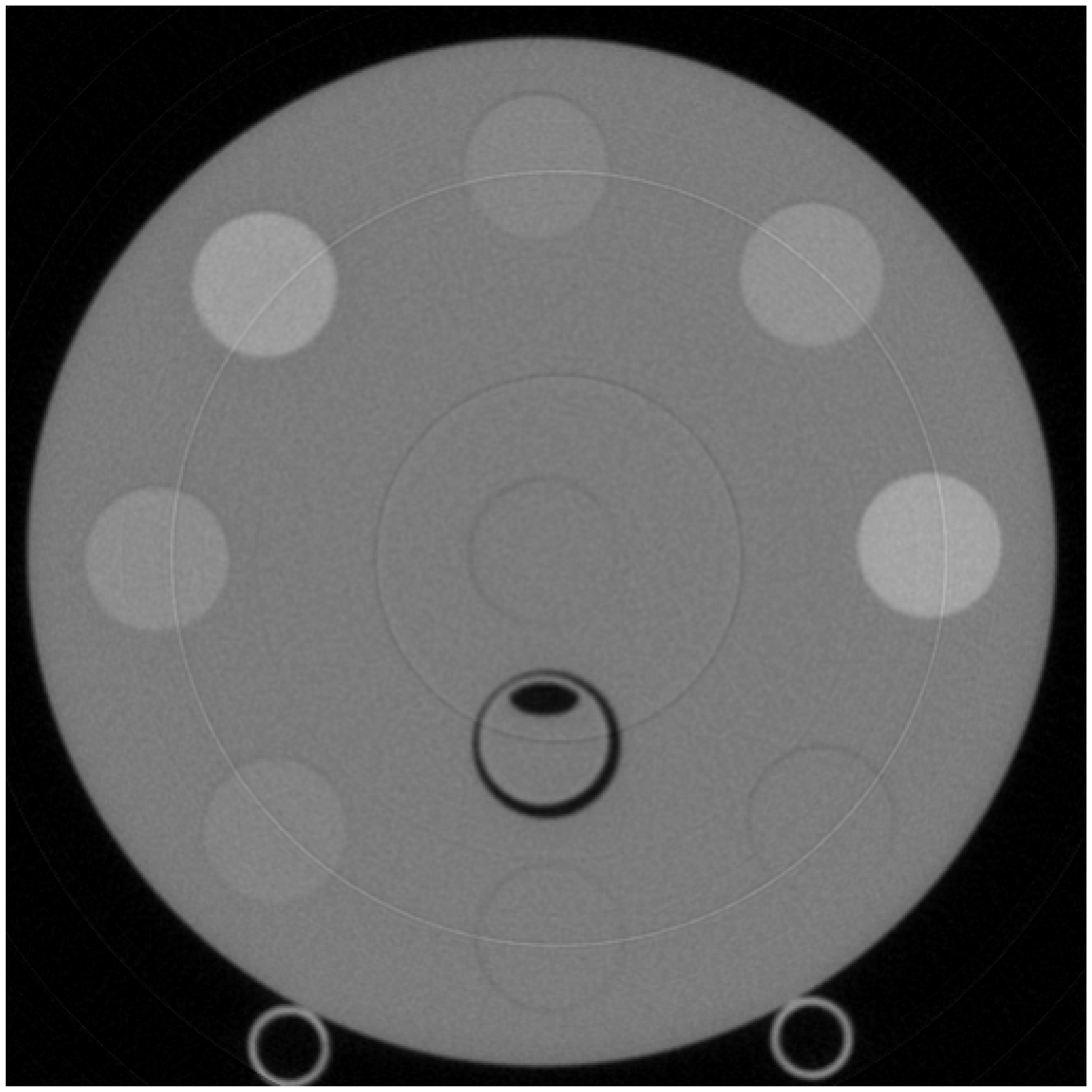}}
         \subcaption{}
         \label{fig:small_gamm_kev_new}
     \end{subfigure}
        \\
\end{tabular}
\caption{MLE and MACE reconstructions from measured data of the Gammex multienergy phantom. Water display window [0, 1450] HU. Iodine display window [-50, 110]~HU. 70~keV display window [0, 2000]~HU.}
\label{fig:small_gamm_mace}
\end{minipage}
\end{figure*}

Figure~\ref{fig:small_gamm_mace} shows a reconstruction, from actual measured data, for the same phantom with the proposed algorithm (bottom row) as well as a comparison to the calculated MLE (top row). The MLE estimates are obtained with $N_{\textrm{MLE}}=100$ to ensure convergence. The proposed method also employs MLE estimate to initialize $p$. However, in this case, we found $N_{\textrm{MLE}}=15$ to be sufficient, saving the computation time.
This demonstrates that we can reduce the number of iterations needed to calculate an approximate value for the MLE while still achieving significant noise reduction. For the prior model to denoise the measured data, we decorrelate the PE and PVC pathlength sinograms by rotating these in material space and applying Gaussian filters of standard deviation 6.0 and 1.5, respectively. We then rotate the pathlength sinograms to their original orientation. Lastly, we clip the values to the limits of the calibration space. Note that this is a more complex denoiser compared to that applied to the simulated data. This demonstrates the modularity of our method with the ability to mix and match denoisers depending on application. Figure~\ref{fig:small_gamm_mace} shows a visual comparison in noise between the MLE and the MACE results.

\begin{figure*}[]
\centering
\begin{minipage}{\textwidth} 
\centering
\begin{tabular}{c c}
\begin{subtable}{0.45\textwidth}
\centering \begin{tabular}{|c|c|c|c|}
\hline
\textbf{Material} & \textbf{Circle} & \textbf{MLE} & \textbf{MACE}\\
\specialrule{.1em}{0em}{0em}
\multirow{4}{*}{Water} & Center  & 964 $\pm$ 175 & 962 $\pm$ 13.0 \\
                \cline{2-4}
                 & Left & 967 $\pm$ 161 & 994 $\pm$ 15.0\\
                 \cline{2-4}
                 & Right & 1254 $\pm$ 197 & 1238 $\pm$ 14.8\\
                 \cline{2-4}
                 & Top & 1037 $\pm$ 163 & 1031 $\pm$ 13.0 \\
\specialrule{.1em}{0em}{0em}
\multirow{4}{*}{Iodine} & Center  & 1.9 $\pm$ 58.4 & 1.9 $\pm$ 2.5 \\
                \cline{2-4}
                 & Left & 47.0 $\pm$ 55.0 & 46.9 $\pm$ 3.5\\
                 \cline{2-4}
                 & Right & 17.9 $\pm$ 65.9 & 21.0 $\pm$ 3.7\\
                 \cline{2-4}
                 & Top & 3.3 $\pm$ 52.3 & 4.7 $\pm$ 2.5\\
\specialrule{.1em}{0em}{0em}
\multirow{4}{*}{70 keV} & Center  & 969 $\pm$ 79.7 & 966 $\pm$ 18.0 \\
                \cline{2-4}
                 & Left & 1098 $\pm$ 73.9 & 1096 $\pm$ 18.2\\
                 \cline{2-4}
                 & Right & 1293 $\pm$ 82.2  & 1284 $\pm$ 17.7\\
                 \cline{2-4}
                 & Top & 1044 $\pm$ 73.8 & 1041 $\pm$ 17.8\\
\hline
\end{tabular}
    \caption{Mean $\pm$ standard deviation}
    \label{tab:small_gamm_stats}
\end{subtable} 
& 
\hspace{1cm}
     \begin{subfigure}[b]{5.7cm}
         \centerline{\includegraphics[width=5.7cm]{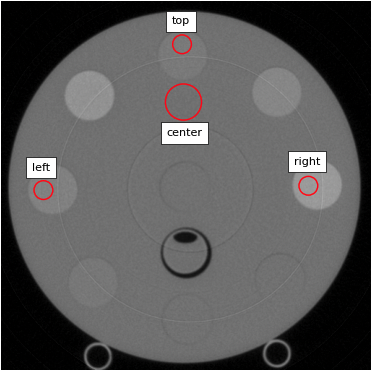}}
         \caption{Figure \ref{fig:small_gamm_kev_new} with representative circles}
         \label{fig:small_gamm_circles}
     \end{subfigure}

\end{tabular}
\caption{Statistics for images in Figure \ref{fig:small_gamm_mace}. Table \ref{tab:small_gamm_stats} shows the mean and standard deviation for each reconstruction and the corresponding circles. The representative circles and their labels appear in Figure \ref{fig:small_gamm_circles}.}
\label{fig:small_gamm_stats_and_circs}
\end{minipage}
\end{figure*}

Figure \ref{fig:small_gamm_circles} labels circles used for quantitative evaluation. Table \ref{tab:small_gamm_stats} shows the means and standard deviations for the selected circles. As expected from Figure~\ref{fig:small_gamm_mace}, our method shows a similar mean but a significant reduction in standard deviation compared to the MLE quantitatively.

Figure~\ref{fig:small_gamm_mace} reconstructions have thin, bright and dark concentric rings, which are artifacts due to limitation of the detector prototype employed. These  measurements are removed in practice, but such removal is not part of this proposed method. Importantly, the proposed method gives flexibility to scale the $\lambda_{\Sigma}$ parameter in a band around these detector elements to remove such artifacts, but the fine tuning of $\lambda_{\Sigma}$ is not performed in these results.

\section{Conclusion}
Our framework provides an efficient reconstruction method to produce CT reconstructions with material decomposition from PCD data.  The modularity of our system is valuable in that it allows for easy modification to better model a wide variety of detectors and system geometry, as well as to make use of advanced prior models, including algorithmic priors such as denoisers or neural networks, either in the sinogram domain or the image domain.  Our results on simulated data indicate an excellent 4.5 times boost in CNR on small low contrast phantom inserts, corresponding to a dose saving of over 20 times. In prototype form, the algorithm requires under 2.5~s of computation per detector row (native slice), which suggests it is practical for implementation in clinical PCCT systems, which are typically deployed with extensive parallel processing resources.

\vspace{-3mm}

\bibliography{main} 
\bibliographystyle{spiebib} 

\end{document}